\documentclass[a4paper,11pt]{article}
\usepackage{jheppub} % for details on the use of the package, please
\usepackage[T1]{fontenc} % if needed

\usepackage{amsmath}
\usepackage{amssymb}
\usepackage{graphicx}
\usepackage{tabularx}
\usepackage{array}
\usepackage{dcolumn}
\usepackage[utf8]{inputenc}

\title{N$_f$ = 1 QCD in  External Magnetic Fields: Staggered Fermions }

\author[a,b]{Paolo Cea}
\author[a]{Leonardo Cosmai}

\affiliation[a]{INFN, Sezione di Bari, I-70126 Bari, Italy}
\affiliation[b]{Dipartimento di Fisica dell'Universit\`a di Bari, I-70126 Bari, Italy}

\emailAdd{paolo.cea@ba.infn.it}
\emailAdd{leonardo.cosmai@ba.infn.it}

\abstract{
We investigate N$_f$ = 1 QCD  in  external magnetic fields on the lattice. The background field is introduced by means of the so-called Schr\"odinger functional. We adopt standard staggered fermions with constant bare mass $am = 0.025$ and  magnetic fields with constant magnetic flux
up to $a^2 e H \simeq   2.3562$. We find that the the deconfinement and chiral symmetry restoration temperatures 
do not depend on the strength of the applied magnetic field.  Our method allow us to easily study the effects of the external magnetic
fields on the QCD thermodynamics. We determine the influences of  applied magnetic fields
to the free energy, pressure, and equation of state  of strongly interacting matter. 
}

\begin{document}
\maketitle
\flushbottom

\section{Introduction}
\label{sec1}
Strong interactions are described by quantum chromodynamics (QCD), a local relativistic non-abelian quantum field theory which is
not amenable to perturbation theory in the low-energy, large-distance regimes.
However, many fundamental questions are linked to the large scale behavior of QCD. In particular,  non-perturbative
approaches to QCD can be used to account for the different phases of hadronic matter under extreme conditions. \\
Recently, the study of the effects of strong magnetic fields on the QCD phase diagram has become a topic of increasing
interest (for a recent review, see Refs.~\cite{Kharzeev:2013jha,Endrodi:2014vza}). In the non-perturbative regimes this problem can be efficiently
approached by lattice QCD simulations with dynamical quarks. \\
The study of lattice gauge theories with  external background fields has been pioneered in Refs.~\cite{Damgaard:1987ec,Damgaard:1988hh} for the U(1) Higgs model in an external electromagnetic field. In the continuum a background field can be introduced by writing:
\begin{equation}
\label{1.1}
A_{\mu}(x) \;  \rightarrow  \; A_{\mu}(x)  \; + \; A^{\text{ext}}_{\mu}(x)  \; .
\end{equation}
In the lattice approach one deals with link variables $U_{\mu}(x)$. Accordingly, on the lattice Eq.~(\ref{1.1}) becomes:
\begin{equation}
\label{1.2}
U_{\mu}(x) \;  \rightarrow  \; U_{\mu}(x) \, U^{\text{ext}}_{\mu}(x)  \; ,
\end{equation}
where $U^{\text{ext}}_{\mu}(x)$ is the lattice version of the background field $A^{\text{ext}}_{\mu}(x)$. As a consequence the
lattice action gets modified as:
\begin{equation}
\label{1.3}
S [U] \;  \rightarrow  \;S [U]   \; +  \delta \, S[U,U^{\text{ext}}] \; ,
\end{equation}
where $ \delta \, S[U,U^{\text{ext}}]$ takes into account the influence of the external field. An alternative method, which is equivalent in the continuum limit,
 is based on the observation that an external background field can be introduced via an external current~\cite{Cea:1989ag,Cea:1990td}:
\begin{equation}
\label{1.4}
J^{\text{ext}}_{\mu}  \;  = \; \partial_{\nu} \;   F^{\text{ext}}_{\nu \mu}  \; ,
\end{equation}
so that the action gets modified as:
\begin{equation}
\label{1.5}
S   \;  \rightarrow  \;S   \; +  \; S_{B} \; , 
\end{equation}
where:
\begin{equation}
\label{1.6}
S_{B}  \; = \;  \int dx  \; J^{\text{ext}}_{\mu} (x)  \; A_{\mu}(x)  \; = \; - \; \frac{1}{2} \;  \int dx  \;  F^{\text{ext}}_{\nu \mu}(x) \,  F_{\nu \mu}(x) \;.
\end{equation}
The main disadvantage of these approaches resides on the lack of gauge invariance  for non-abelian gauge theories.
The issue of gauge invariance, however, does not pose if one is interested in QCD in external magnetic fields. In fact,
let us consider the lattice partition function of QCD with $f$ flavors of dynamical staggered quarks:
\begin{equation}
\label{1.7}
 Z  \;  = \; \int {\mathcal{D}}U \; e^{-S_G} \; \prod_f  \left [  det M(U)   \right ]^{\frac{1}{4}} \; ,
\end{equation}
where $S_G$ is the gauge field action and $M$ is the fermion matrix for a staggered quark with bare mass $am_f$:
\begin{eqnarray}
\label{1.8}
&& M_{n,m}(U) = \sum_{\nu =1}^{4}  \frac{\eta_{\nu}(n)}{2} \left\{ U_{\nu}(n)
      \delta_{m,n+\hat{\nu}}-
    U_{\nu}^{\dagger}(m)  \delta_{m,n-\hat{\nu}} \right\}  \, + \, am_f  \, \delta_{m,n} \; ,\nonumber  \\    
&&   \; \; \; \;  \eta_{\nu}(n) = (-1)^{n_1+\ldots+n_{\nu-1}} \; .
\end{eqnarray}
Since magnetic fields couple only to quarks, external magnetic fields can be introduced in the lattice action by replacing in the fermion
mass matrix  Eq.~(\ref{1.8}) the gauge field links  according to  Eq.~(\ref{1.2}), where  the $U^{\text{ext}}_{\mu}(x)$'s are U(1) elements
corresponding to the external magnetic fields with continuum gauge potential  $A^{\text{ext}}_{\mu}(x)$. For instance, if we consider constant
magnetic fields directed along the $x_3$ direction, then the continuum  gauge potential in the Landau gauge reads:
\begin{equation}
\label{1.9}
A^{\text{ext}}_k(\vec{x}) =  \delta_{k,2} \ x_1 H \; .
\end{equation}
Therefore, we may write:
\begin{equation}
\label{1.10}
U^{\text{ext}}_1(\vec{x}) = U^{\text{ext}}_3(\vec{x}) = U^{\text{ext}}_4(\vec{x}) = 1 \, , \;\;\;
\,  U^{\text{ext}}_2(\vec{x}) = \cos(  q_f eH x_1) +
i \sin( q_f eHx_1 )   \; ,
\end{equation}
where  $e$ is the (positive) elementary charge and $q_f$ is the quark charge ($q_u = 2/3$ , $q_d = - 1/3$).
 Since the lattices  have  the topology of a torus, the magnetic field turns out to be quantized~\cite{Damgaard:1988hh}:
\begin{equation}
\label{1.11}
  a^2 q_f \, e H \; = \; \frac{2 \pi}{L_s^2} \; 
n_{\text{ext}} \; , \;\;\;  n_{\text{ext}}\,\,\,{\text{integer}} 
\end{equation}
where $L_s$ is the lattice spatial size. Indeed, in the recent literature this approach has been adopted in extensive numerical
simulations of QCD in external magnetic fields~\cite{D'Elia:2010nq,D'Elia:2011zu,Braguta:2011hq,Ilgenfritz:2012fw,Bali:2011qj,Bali:2012jv,Bali:2013esa,Bonati:2013lca,D'Elia:2012zw,Levkova:2013qda,Ilgenfritz:2013ara,Bali:2013owa,Bonati:2013vba,Bornyakov:2013eya,Bali:2014kia,Borsanyi:2013bia,Endrodi:2015oba}. \\
An alterative approach to put background fields on the lattice has been proposed  since long time~\cite{Cea:1996ff}.
Indeed, that proposal allows to overcome the problem of gauge invariance by implementing  background fields on the lattice  by means of 
the manifestly gauge-invariant lattice Schr\"odinger functional. In this paper we present an exploratory study of lattice QCD 
immersed in a uniform external magnetic field. The background field is introduced by using the  Schr\"odinger functional.
Moreover, for simplicity,  we restrict ourself to one flavor of staggered dynamical quark. \\
The plan of the paper is as follows. In Sect.~\ref{sec2}, for completeness,
we briefly discuss our method to introduce  background fields on the lattice. In Sect.~\ref{sec3} 
we present the results of our  numerical simulations for several local observables. We also address the problem of the
possible dependence of the pseudoritical couplings on the magnetic field strengths.
Sect.~\ref{sec4}  is devoted to the discussion of the effects of magnetic fields on QCD  thermodynamics.
Finally, our conclusions are relegated in  Sect.~\ref{sec5}.
\section{Magnetic Fields within the Schr\"odinger Functional}
\label{sec2}
For reader's convenience, let us briefly review  background fields in lattice gauge theories
within the Schr\"odinger functional. Firstly, we illustrate the method in pure gauge theories.
In Ref.~\cite{Cea:1996ff}, to overcome the gauge invariance  problem  in presence of background fields, it was
proposed that  background fields on the lattice could be implemented by means of 
the gauge invariant lattice Schr\"odinger functional:
\begin{equation}
\label{2.1}
 {\mathcal{Z}}[U^{\mathrm{ext}}_k] = \int {\mathcal{D}}U \; e^{-S_G}  \; ,
\end{equation}
where the functional integration is extended over links on a lattice with the
hypertorus geometry  and satisfying the constraints ($x_t \equiv x_4$ is the  temporal coordinate)
\begin{equation}
\label{2.2}
U_k(x)|_{x_t=0} = U^{\mathrm{ext}}_k(\vec{x}) \,, \; \; (k=1,2,3).
\end{equation}
One also imposes that links at the spatial boundaries are fixed according to Eq.~(\ref{2.2}). In fact, in the continuum this last
condition amounts to the requirement that fluctuations over the background field vanish at infinity. 
This approach has been applied for both abelian and non-abelian gauge theories with different background 
fields~\cite{Cea:1996rq,Cea:1996sw,Cea:1999gn,Cea:2000zr,Cea:2000xi,Cea:2001an,Cea:2001ef,Cea:2002wx,Cea:2005td,Cea:2005dg}.
\\
The effects of dynamical fermions can be accounted for quite easily. Indeed, when including dynamical fermions, the lattice Schr\"odinger functional
in presence of a static external background gauge field becomes:
\begin{eqnarray}
\label{2.3}
\mathcal{Z}[U^{\mathrm{ext}}_k]  &=& 
\int_{U_k(L_t,\vec{x})=U_k(0,\vec{x})=U^{\text{ext}}_k(\vec{x})}
\mathcal{D}U \,  {\mathcal{D}} \psi  \, {\mathcal{D}} \bar{\psi} e^{-(S_G+S_F)} 
\nonumber \\ 
&=&  \int_{U_k(L_t,\vec{x})=U_k(0,\vec{x})=U^{\text{ext}}_k(\vec{x})}
\mathcal{D}U e^{-S_G} \, \det M \,,
\end{eqnarray}
where $S_F$ is the fermion action and $M$ indicates   the generic fermion matrix.
Notice that the fermion fields are not constrained and
the integration constraint is only relative to the gauge fields.
This leads  to the appearance of  the gauge invariant fermion determinant after integration on the 
fermion fields.  As usual we impose on fermion fields
periodic boundary conditions in the spatial directions and
anti-periodic boundary conditions in the temporal direction.  In fact, Eq.~(\ref{2.3}) has been employed to study
the dynamics of QCD with two degenerate staggered quarks~\cite{Cea:2004ux,Cea:2007yv},
as well as the quantum Hall effect in graphene~\cite{Cea:2012up}. \\
In the case of QCD in constant magnetic fields the  constraints in the lattice Schr\"odinger functional need to be slightly modified
to take into account that the magnetic field is coupled only to quarks. To this end, we impose that during the upgrade of the gauge links
$U^{\mathrm{ext}}_k(\vec{x}) =  \mathbb{I}$, while for the upgrade of the fermion fields  
$U^{\mathrm{ext}}_k(\vec{x}) =  \mathbb{I} \times e^{i  \theta^{\mathrm{ext}}_{k}(\vec{x})}$ where:
\begin{equation}
\label{2.4}
 \theta^{\text{ext}}_k(\vec{x})  \; = \;  \delta_{k,2}  \;   q_f \, eH \, x_1 \; .
\end{equation}
Since our Schr\"odinger functional ${\mathcal{Z}}[U_k^{\mathrm{ext}}]$ is defined on a lattice
with periodic boundary conditions, usually we impose that:
\begin{equation}
\label{2.5}
\theta_2(x_1,x_2,x_3,x_4)  \; =  \; \theta_2(x_1+L_s ,x_2,x_3,x_4) \; .
\end{equation}
As a consequence the magnetic field $H$ turns out to be quantized:
\begin{equation}
\label{2.6}
a^2  q_f  \, eH  \; = \; \frac{2 \pi}{L_s} \;  n_{\mathrm{ext}} 
\end{equation}
with $n_{\mathrm{ext}}$ integer. However, it should be kept in mind that we are dealing with a periodic lattice with
fixed boundary conditions, so that is is not strictly necessary to impose the quantization Eq.~(\ref{2.6}) and the
"integer"  $n_{\mathrm{ext}}$ can be an  arbitrary  real number. 
\section{Numerical Results}
\label{sec3}
We perform simulations of lattice QCD with one-flavor of rooted staggered quark.
Our numerical results were obtained by choosing as gauge action  the Wilson action:
\begin{equation}
\label{3.1}
S_G \;  = \; \beta \; S_W  \equiv  \beta  \sum_{x,\mu>\nu}  \,
\left ( 1 - \frac{1}{3} \,  {\mathrm{Re}} \; [{\mathrm {Tr} } \, U_{\mu\nu}(x)] \right )
\end{equation}
where $U_{\mu\nu}(x)$ are the plaquettes in the $(\mu,\nu)$-plane and   $\beta=\frac{6}{g^2 }$.
Therefore we are led to consider the following lattice  Schr\"odinger functional: 
\begin{equation}
\label{3.2}
 \mathcal{Z}[U^{\mathrm{ext}}_k]  \;  = \; \int_{U_k(L_t,\vec{x})=U_k(0,\vec{x})=U^{\text{ext}}_k(\vec{x})}
   {\mathcal{D}}U \; e^{- \beta S_W} \;   \left [  {\mathrm{det}} \,  M(U)   \right ]^{\frac{1}{4}} \; ,
\end{equation}
where the staggered fermion matrix is given by  Eq.~(\ref{1.8}).
To perform the functional integration over the  SU(3) links we have made use of  the publicly available MILC code~\cite{milc}  
which has been suitably modified by us in order to introduce the boundary constraints  Eq.~(\ref{2.2}). 
All simulations make use of the rational hybrid Monte Carlo (RHMC) algorithm. 
The functional integration is performed over the lattice links,
but  the links at the spatial boundaries are fixed according to Eq.~(\ref{2.2}). 
Accordingly, the links which are frozen are not evolved during the molecular
dynamics trajectory and the corresponding conjugate momenta are set to zero.
The length of each RHMC trajectory has been set to 1.0 in molecular dynamics time units.
For each value of the gauge coupling $\beta$ and the magnetic field $eH$ we collected  4000 - 5000
trajectories, and about 10000 trajectories around the critical coupling.
To allow thermalization we typically discarded 1000 trajectories. 
The statistical errors were estimated by means of boostrap  combined with binning. \\
In the present exploratory study we consider lattices of size $L_s = 24$ and $L_t = 4$ and  fixed  bare fermion mass 
$m_0 \equiv  am = 0.025$. At fixed $L_t$ the temperature of the gauge system $T = \frac{1}{a L_t }$ is 
changed by varying the coupling constant $\beta$.  \\
Since the smallest quark electric charge is $|q| = 1/3$, from   Eq.~(\ref{2.6}) we get: 
\begin{equation}
\label{3.3}
a^2  \, eH  \; = \; \frac{6 \pi}{L_s} \;  n_{\mathrm{ext}} 
\end{equation}
Different strengths of the external magnetic field are labelled by the parameter $n_{\text{ext}}$ according to Eq.~(\ref{3.3}).
We performed simulations for  $n_{\text{ext}} = \, 0 \, , \, 1 \, , \, 3$, corresponding to field strength  
$a^2 e H =  \, 0 \, , \, 0.7854 \, , \,   2.3562$ in lattice units,
and assumed $q_f = \frac{2}{3}$ (up quark). Note that, the case of down quark  $q_f = - \frac{1}{3}$
can be recovered with  $n_{\text{ext}} = -\, \frac{1}{2} $ ($a^2 e H = \, - \, 0.3927$). 
In fact, to check the dependence of the free energy on the magnetic field we have also performed numerical simulations 
for $n_{\text{ext}} = -\, \frac{1}{2} $ .

For the sake of completeness let us discuss, briefly, how the background magnetic field influences the dynamics of the gauge system. We said that to update the gauge system we used the rational hybrid Monte Carlo algorithm. As it is well known (see for instance Ref.~\cite{degrand2006lattice}), to simulate the fermion determinant one introduces color-triplet scalar pseudofermion fields. The pseudofermion action depends on the inverse of the staggered fermion matrix
Eq.~(\ref{1.8}). In the molecular dynamics one solves the equations of motion of the momenta conjugated to the gauge links. The derivative of the gauge momentum is called the force term, which is the formal derivative of the effective action with respect to the gauge potential. Thus, the force term consists of two contributions, namely the gauge force term and the fermion force term. Our boundary conditions correspond to set $U_k^{\text{ext}}=\mathbb{I}$ on the $x_t=0$ hypersurface and at the  spatial boundaries of the lattice in the gauge force term, while 
$U^{\mathrm{ext}}_k(\vec{x}) =  \mathbb{I} \times e^{i  \theta^{\mathrm{ext}}_{k}(\vec{x})}$ in the fermion force term.
To maintain the above constraints during the molecular dynamics, the momenta conjugated to the frozen gauge links are set to zero.
\subsection{Local Observables}
\label{sec3.1}
\begin{figure}[h] 
\centering
\includegraphics[width=0.7\textwidth]{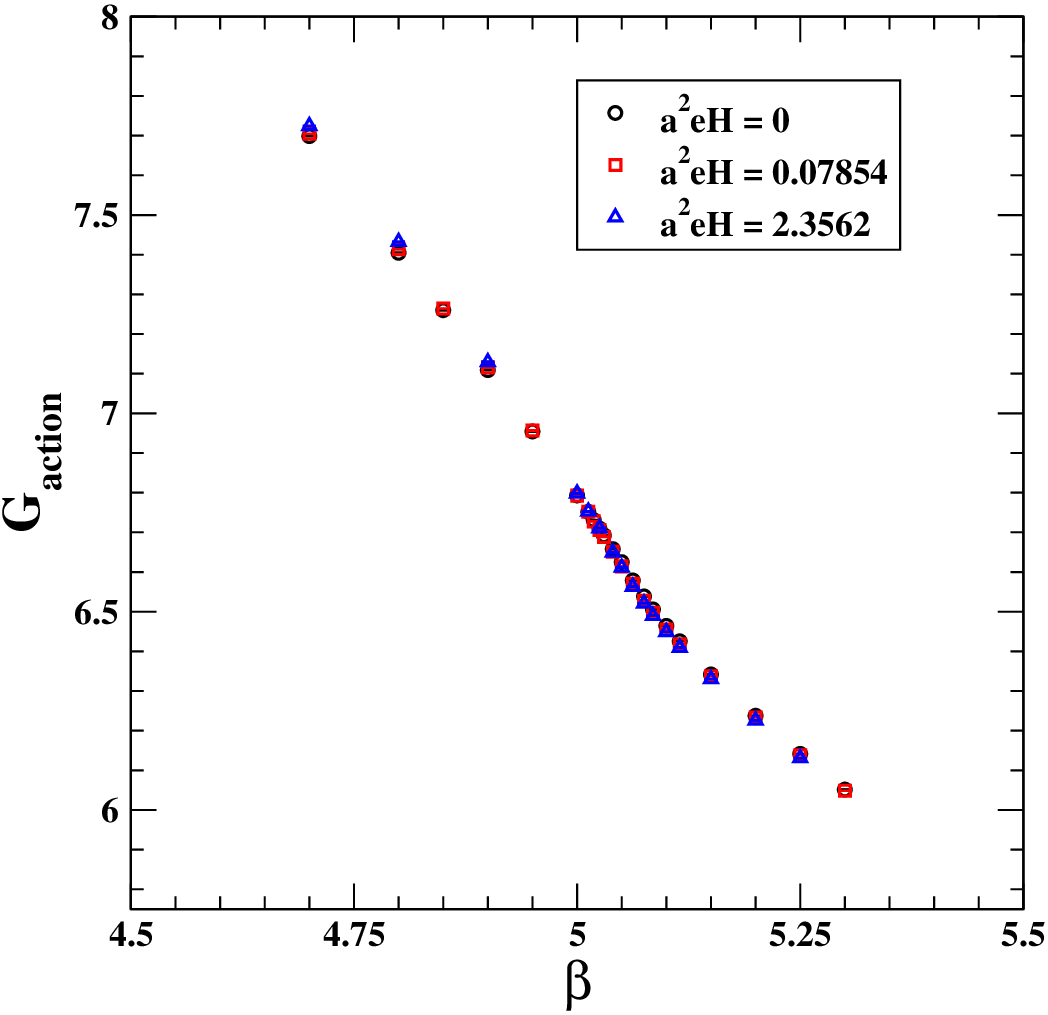} 
\caption{The gauge action Eq.~(\protect\ref{3.4}) versus $\beta$
for   $q_f = \frac{2}{3}$  and   $n_{\text{ext}} =  \, 0 \, , \, 1 \, , \, 3$.}
\label{gaction}
\end{figure}
In this Section we are interested in the effects of the external magnetic field on several local observables.
First, we consider the gauge action which, following the MILC convention,  we define as:
\begin{equation}
\label{3.4}
G_{action} \;  = \; \frac{1}{L_s^3 \, L_t} \;  \left  \langle  \sum_{x,\mu>\nu}  \,
\left [ 3 -  \,  {\mathrm{Re}} \;  {\mathrm{Tr}} \, U_{\mu\nu}(x)  \right ] \right \rangle \; .
\end{equation}
In Fig.~\ref{gaction} we display the gauge action as a function on the gauge coupling $\beta$ for three different values
of the magnetic field. Since $ G_{action}$ is a pure gauge quantity, it couples to the magnetic field only through
quark loops. Therefore we expect that this quantity should manifest a very weak dependence on the magnetic field.
Indeed,   Fig.~\ref{gaction} shows that the effects of the magnetic field on the gauge action are at most of  order $10^{-2}$ 
(see Fig.~\ref{delta_gaction}). Interestingly enough, we see that the gauge action increases as a function of $eH$ in the strong
coupling region whereas it decreases in the weak coupling region. In fact the three different curves displayed in 
Fig.~\ref{gaction} cross near the critical coupling $\beta_c \simeq 5.0$. \\
\begin{figure}[h] 
\centering
\includegraphics[width=0.7\textwidth]{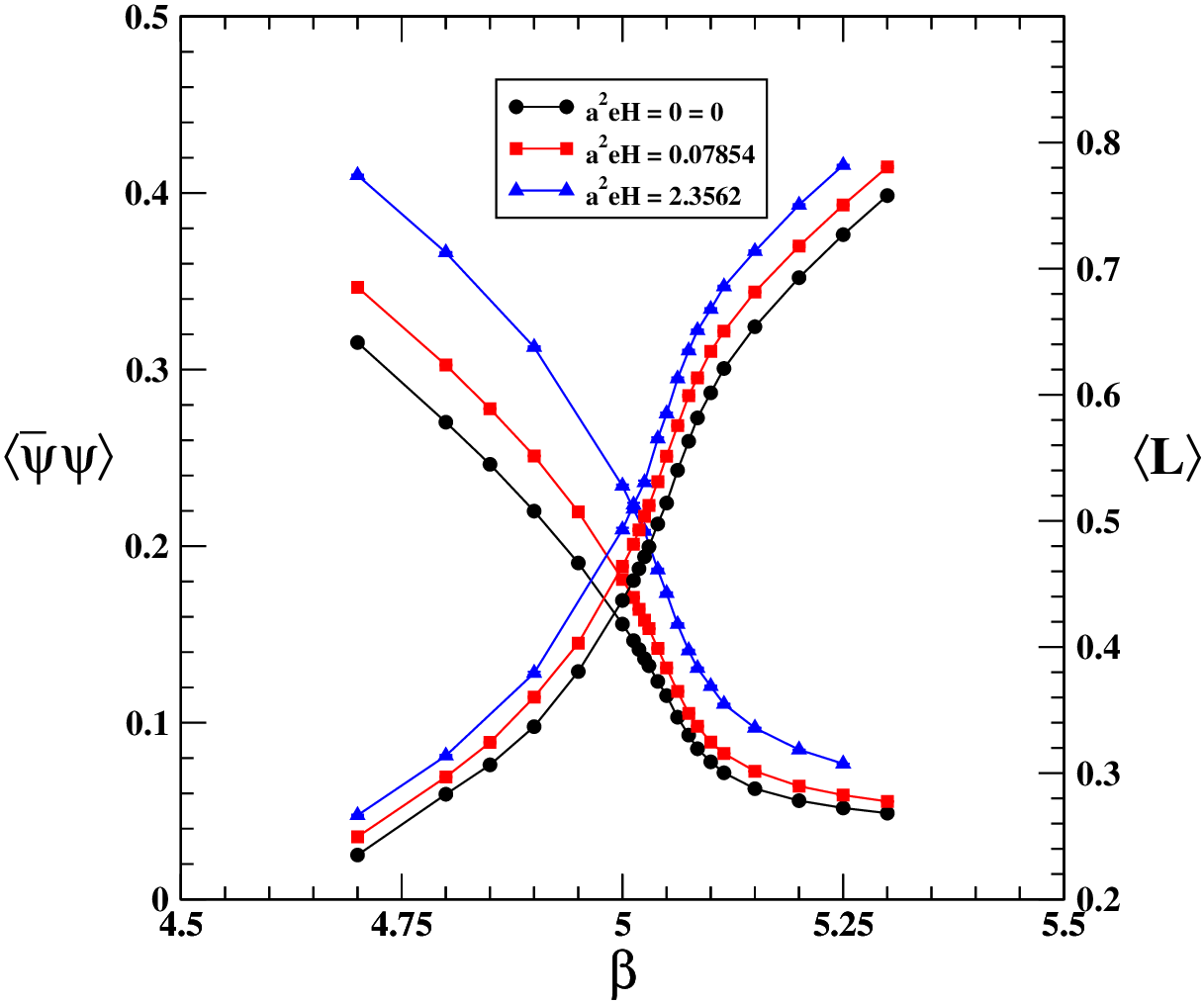} 
\caption{The chiral condensate Eq.~(\protect\ref{3.5}) (left) and
the real value of the Polyakov loop  Eq.~(\protect\ref{3.6}) (right) 
 versus $\beta$  for   $q_f = \frac{2}{3}$  and   $n_{\text{ext}} =  \, 0 \, , \, 1 \, , \, 3$.}
\label{chiralcond_polyakov}
\end{figure}
A more interesting quantity is the quark chiral condensate:
\begin{equation}
\label{3.5}
\langle \overline\psi \, \psi \rangle  \;  = \; \frac{1}{ L_s^3 \, L_t} \;  \frac{1}{4} \,  \left  \langle  
 {\mathrm Tr}  \; M^{-1} \right \rangle \; ,
\end{equation}
which should display a pronunciate dependence on the magnetic field.
In  Fig.~\ref{chiralcond_polyakov}  we display  $\langle \overline\psi \, \psi \rangle$ versus the gauge coupling $\beta$
for three different values of the magnetic field $eH$. The chiral condensate   was computed by noise estimators with 4 random vectors.
It is evident that the chiral condensate increases as a function of $eH$ for all temperatures. For comparison,
in  Fig.~\ref{chiralcond_polyakov}  we also display  the real part of the Polyakov loop expectation value:
\begin{equation}
\label{3.6}
\langle L \rangle  \;  = \; \frac{1}{3 \, L_s^3 } \;  \left  \langle  \sum_{\vec{x}}  \prod_{x_t = 0}^{L_t}  
\;  {\mathrm Tr} \; U_{4}(\vec{x},x_t)   \right \rangle \; .
\end{equation}
The Polyakov loop $L$, likewise the gauge action, is a pure gauge observable. Nevertheless,  Fig.~\ref{chiralcond_polyakov} 
shows that the  Polyakov loop displays a sizable dependence on the magnetic field. In particular, we see that $L$ increases
with $eH$ for all temperatures  as for the chiral condensate. This behavior can be qualitatively understood if the quark free energy
decreases with the applied magnetic field. In fact, later on we will show that the strongly interacting  system behaves  like
a paramagnetic medium, i.e.  positive   magnetic susceptibility. \\
Another interesting feature of   Fig.~\ref{chiralcond_polyakov}  is the crossing of the chiral condensate and the Polyakov loop near
the critical temperature. In fact,   Fig.~\ref{chiralcond_polyakov}  seems to suggest that the pseudocritical gauge coupling $\beta_c$
does not manifest a strong dependence on the magnetic field $eH$. 

\subsection{Pseudocritical couplings}
\label{sec3.2}
\begin{figure}[t] 
\centering
\includegraphics[width=0.7\textwidth]{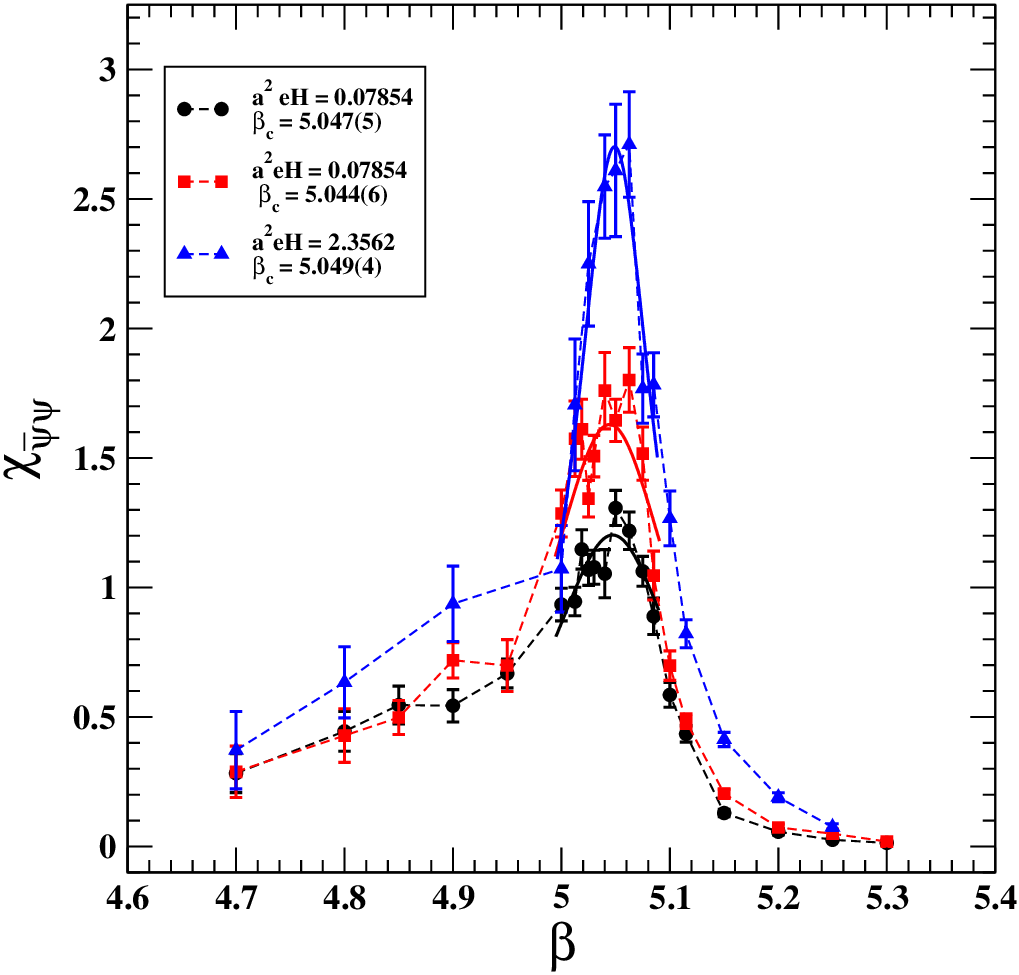} 
\caption{The  disconnected chiral susceptibility  Eq.~(\protect\ref{3.8})
 versus $\beta$  for  $q_f = \frac{2}{3}$  and   $n_{\text{ext}} =  \, 0 \, , \, 1 \, , \, 3$.
 The  continuous lines are the results of the fits of the chiral susceptibilities to Eq.~(\protect\ref{3.7}).
 The estimate critical couplings are reported in the legend. }
\label{chiralcond_suscep}
\end{figure}
In this section we address the problem of the possible dependence of the pseudocritical coupling on the magnetic field.
In general, the (pseudo)critical coupling  is  determined as the value for which some relevant susceptibilities exhibit a peak. 
In the present paper,  to precisely localize the peak in  the relevant susceptivity we  parametrize
the peak region with  a Lorentzian function:
\begin{equation}
\label{3.7}
F(\beta) \;  = \; \frac{a_1}{a_2 \; (\beta \, - \beta_c)^2 \; + \; 1}  \;   \; . 
\end{equation}
Our estimate of the critical coupling $\beta_c$ is obtained by fitting the susceptivity to Eq.~(\ref{3.7}) in the peak region. 
We use the Polyakov loop susceptibility as well as the disconnected part of the chiral susceptibility to locate the
transition temperature to the high temperature phase of QCD. \\
First, we consider the  disconnected  chiral susceptibility:
\begin{equation}
\label{3.8}
\chi^{disc}_{\overline\psi \, \psi}  \;  = \; \frac{1}{ L_s^3 \, L_t} \;  \frac{1}{16} \,  \left (  
 \langle  [{\mathrm Tr}  \; M^{-1}]^2   \rangle  \; - \;  [ \langle  {\mathrm Tr}  \; M^{-1}  \rangle ]^2 \right ) \; .  
\end{equation}
In Fig.~\ref{chiralcond_suscep} we show  the disconnected  chiral susceptibility as a function of the gauge coupling
for three different values of $eH$. 
\begin{figure}[t] 
\centering
\includegraphics[width=0.7\textwidth]{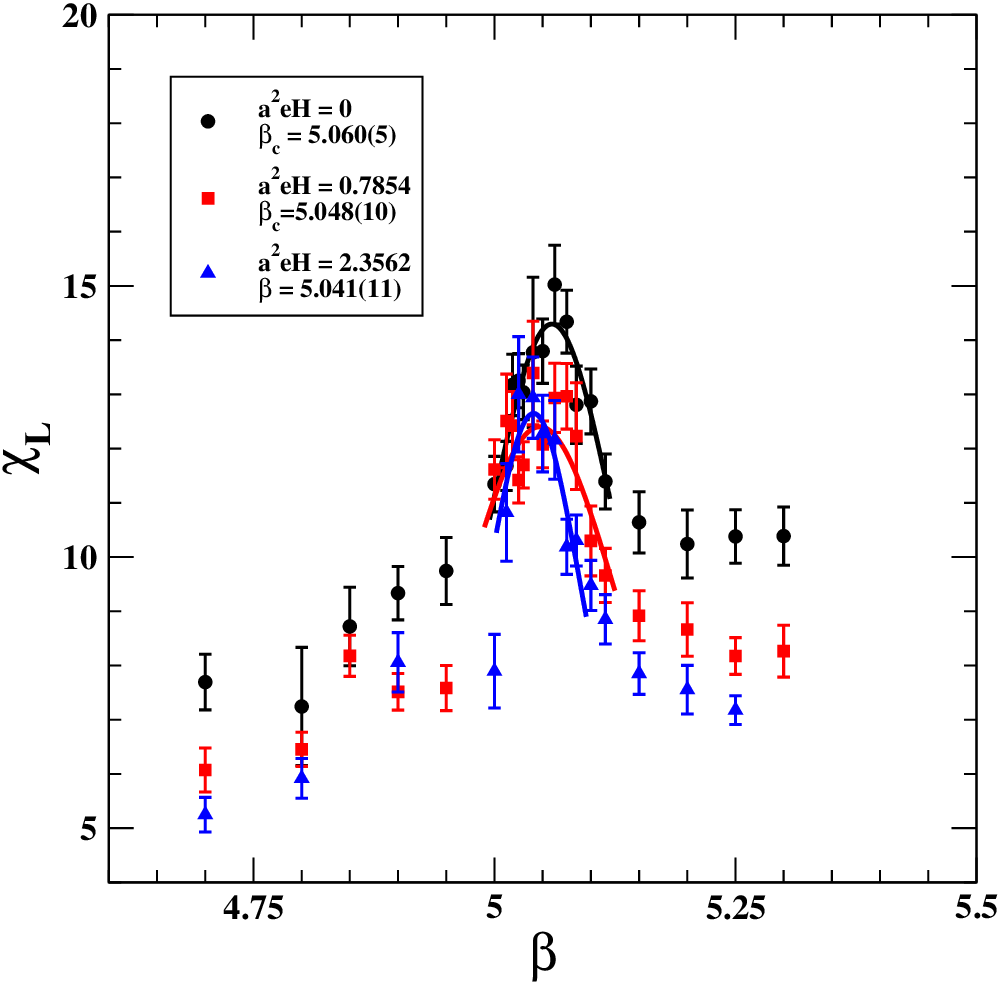} 
\caption{The  Polyakov loop susceptibility  Eq.~(\protect\ref{3.9})
 versus $\beta$  for  $q_f = \frac{2}{3}$  and   $n_{\text{ext}} =  \, 0 \, , \, 1 \, , \, 3$.
 The  continuous lines are the results of the fits of the Polyakov loop susceptibilities to Eq.~(\protect\ref{3.7}).
 The estimate critical couplings are reported in the legend.}
\label{polyakov_suscep}
\end{figure}
\begin{figure}[h] 
\centering
\includegraphics[width=0.7\textwidth]{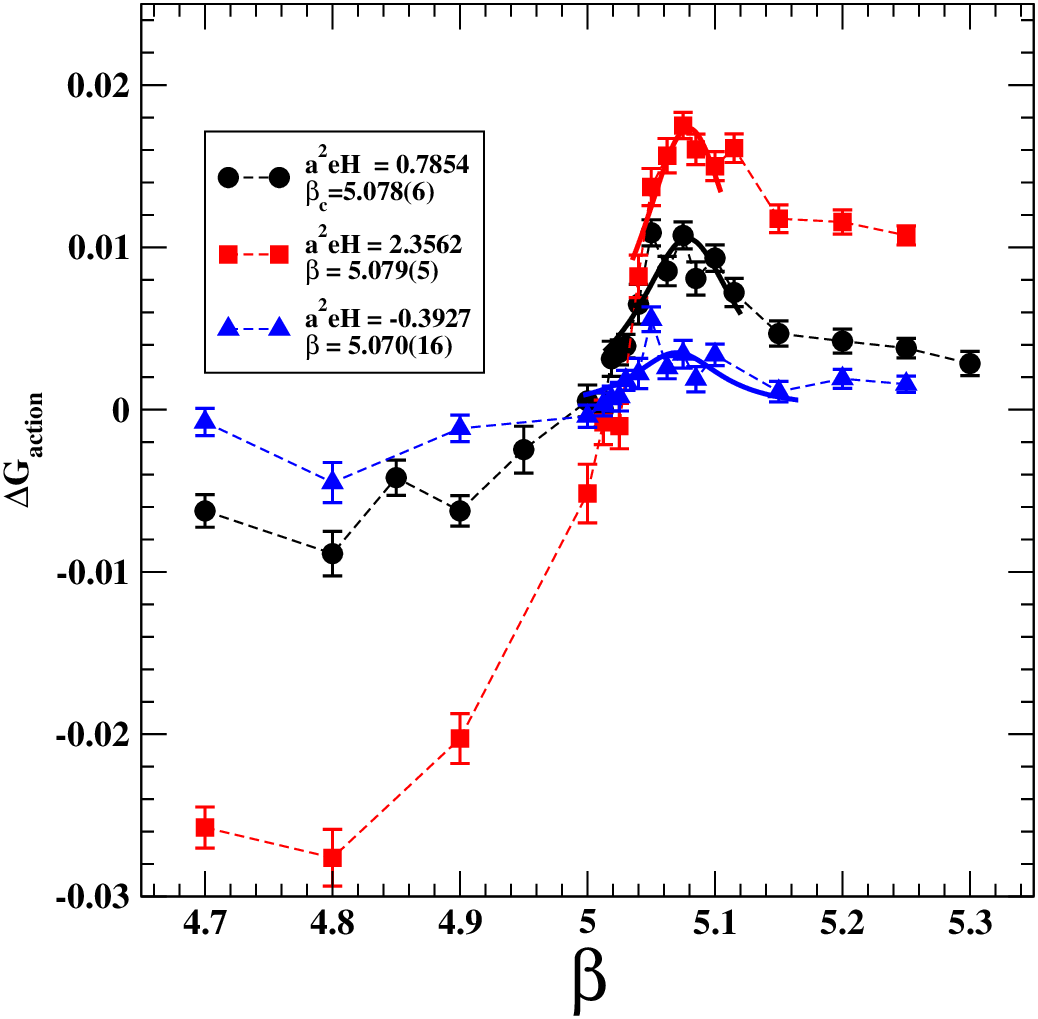} 
\caption{ $\Delta G_{action}$   Eq.~(\protect\ref{3.9})
 versus $\beta$  for  $q_f = \frac{2}{3}$  $n_{\text{ext}} =  \, 1 \, , \, 3$, 
 and   $q_f =  - \frac{1}{3}$   $n_{\text{ext}} =  \, 1$.
The  continuous lines are the results of the fits of $\Delta G_{action}$ to Eq.~(\protect\ref{3.7}).
 The estimate critical couplings are reported in the legend.}
\label{delta_gaction}
\end{figure}
As usual the disconnected chiral  susceptibility displays a sharp peak near the chiral
critical coupling. Interestingly enough, the dependence of the chiral  susceptibility on the magnetic field
is almost relegated to the peak region. Moreover,  the peak values increase with $eH$ signaling 
that the chiral transition sharpens in presence of a non-zero magnetic field.  Notwithstanding, 
we find that, within our statistical uncertainties,
the chiral critical coupling does not depend on the magnetic field. \\
We have also considered the Polyakov loop susceptibility:
\begin{equation}
\label{3.9}
\chi_L \;  = \; L_s^3  \;  \left (  \langle  L^2   \rangle \;  - \;  \langle  L  \rangle^2    \right ) \; . 
\end{equation}
Results for the Polyakov loop susceptibility are shown in Fig.~\ref{polyakov_suscep}. In this case we see that 
the dependence of the  Polyakov loop susceptibility on the magnetic field is less pronounced   with respect
to the  disconnected chiral  susceptibility. Moreover,  Fig.~\ref{polyakov_suscep} shows that the peak
of  $\chi_L$  decreases with increasing $eH$.  This means that the applied magnetic field tends to smooth out the
deconfinement transition. However, even in this case the deconfinement critical couplings does not display
any appreciable dependence on $eH$. Moreover, we find that the chiral and deconfinemnet critical couplings 
agree for any values of the magnetic field  strengths  considered in this work.
\begin{table}[thb]
\setlength{\tabcolsep}{1pc}
\centering
\caption[]{Summary of the values of the critical couplings $\beta_c$ estimated
 with  different operators for  the magnetic field  strenghts  considered in this work.} 
\begin{tabular}{cll}
\hline
\hline
Operator & $a^2 eH$ & $\beta_c$   \\
\hline
$\chi^{disc}_{\overline\psi \, \psi} $  & ~ 0      & 5.047(5)  \\
                 & ~  0.7854    &   5.044(6)  \\
                &  ~ 2.3562     &  5.049(4)  \\
                &  $-$0.3927 &  5.046(2) \\
 \hline
$\chi_L$   & ~ 0      & 5.060(5)  \\
                 & ~ 0.7854    &   5.048(10)  \\
                &  ~ 2.3562     &   5.041(11)    \\
                & $-$0.3927  &   5.051(5)   \\
\hline
$\Delta G_{action}$  & ~ 0      & --  \\
                 &  ~ 0.7854    &  5.078(6)  \\
                & ~ 2.3562     &  5.079(5)  \\
                & $-$0.3927  &  5.070(16) \\

\hline
\hline
\end{tabular}
\label{summary}
\end{table}
Finally, as further check, we have considered  the variation of the gauge action:
\begin{equation}
\label{3.10}
\Delta G_{action} \;  \equiv  \;   G_{action}(eH \neq 0) \; -  G_{action}(eH=0) \; ,
\end{equation}
which is known to  display a peak in the critical region.  Of course, by using $ \Delta G_{action}$ we may 
estimate the pseudocritical couplings only for non-zero magnetic field strengths. \\
In Fig.~\ref{delta_gaction} we show   $ \Delta G_{action}$ versus $\beta$ for the  three different values of the
magnetic field employed in the present work.  As discussed in Sec.~\ref{sec3.1} the effects of the magnetic field on the gauge action
are tiny. Moreover, the non monotonic dependence of the gauge action on $eH$ is clearly displayed
in  Fig.~\ref{delta_gaction}. In any case, we see that  $\Delta G_{action}$  does display a well
developed peak in the critical region.
We find that the peaks in  $\Delta G_{action}$ are located at a
systematically slightly larger values of the gauge coupling with respect to the chiral and Polyakov loop
susceptibilities.  In a finite volume this is, of course, not unexpected. Indeed,
we recall that we are using  lattices with fixed boundary conditions and, as  previous studies showed, 
the gauge action turns out to be more susceptible to finite volume effects. 
Nonetheless, what it is relevant is that the pseudocritical couplings do not depend on the magnetic field $eH$. \\
For reader convenience, in Table~\ref{summary} we summarize our estimates of the critical couplings $\beta_c$ 
as function of the magnetic field $eH$. From  Table~\ref{summary} we may safely conclude that the critical temperature
does not depend on the external magnetic field.
\section{Thermodynamics in external Magnetic Fields}
\label{sec4}
The partition function $\mathcal{Z}[U^{\mathrm{ext}}_k]$ allows us to define observables that can be used to establish the
equation of state of the theory. Such observables play an important role in describing the thermodynamic
properties of the system. \\
The free energy density is related to the logarithm of the partition function as:
\begin{equation}
\label{4.1}
f(T,H) \; = \;  - \; \frac{T}{V} \; \log  \mathcal{Z}[U^{\mathrm{ext}}_k] \; = \;
- \;   \frac{1}{L_s^3 \, L_t} \; \log  \mathcal{Z}[U^{\mathrm{ext}}_k]  \; .
\end{equation}
The pressure is given by the derivative of  $T  \log  \mathcal{Z}[U^{\mathrm{ext}}_k]$  with respect to the volume. Assuming that we
have a large, homogeneous system, differentiation with respect to V is equivalent to dividing
by the volume. Therefore in the thermodynamic limit the pressure can be written as minus
the free energy density:
\begin{equation}
\label{4.2}
p(T,H) \; = \; - \; f(T,H) \; .
\end{equation}
Using the well-known relation between  the  trace anomaly (also called interaction measure) and the derivative of the pressure:
\begin{equation}
\label{4.3}
I(T,H) \; = \;  \; \varepsilon(T,H) \, - \, 3 \, p(T,H) \; =  \;  T^5 \;  \frac{\partial}{\partial \, T} \; \frac{p(T,H)}{T^4} \; ,
\end{equation}
one can easily calculate the energy density: 
\begin{equation}
\label{4.4}
\varepsilon(T,H)  \; = \;  I(T,H) \;  +  \;  3 \, p(T,H) \; ,
\end{equation}
the entropy density:
\begin{equation}
\label{4.5}
s(T,H)  \; = \;  \frac{ \varepsilon(T,H)  +  \;  p(T,H)}{ T} \; ,
\end{equation}
and the speed of sound:
\begin{equation}
\label{4.6}
c^2_s  \; = \;  \left .  \frac{\partial p}{\partial \varepsilon} \right  |_s \; .
\end{equation}
\begin{figure}[th] 
\centering
\includegraphics[width=0.7\textwidth]{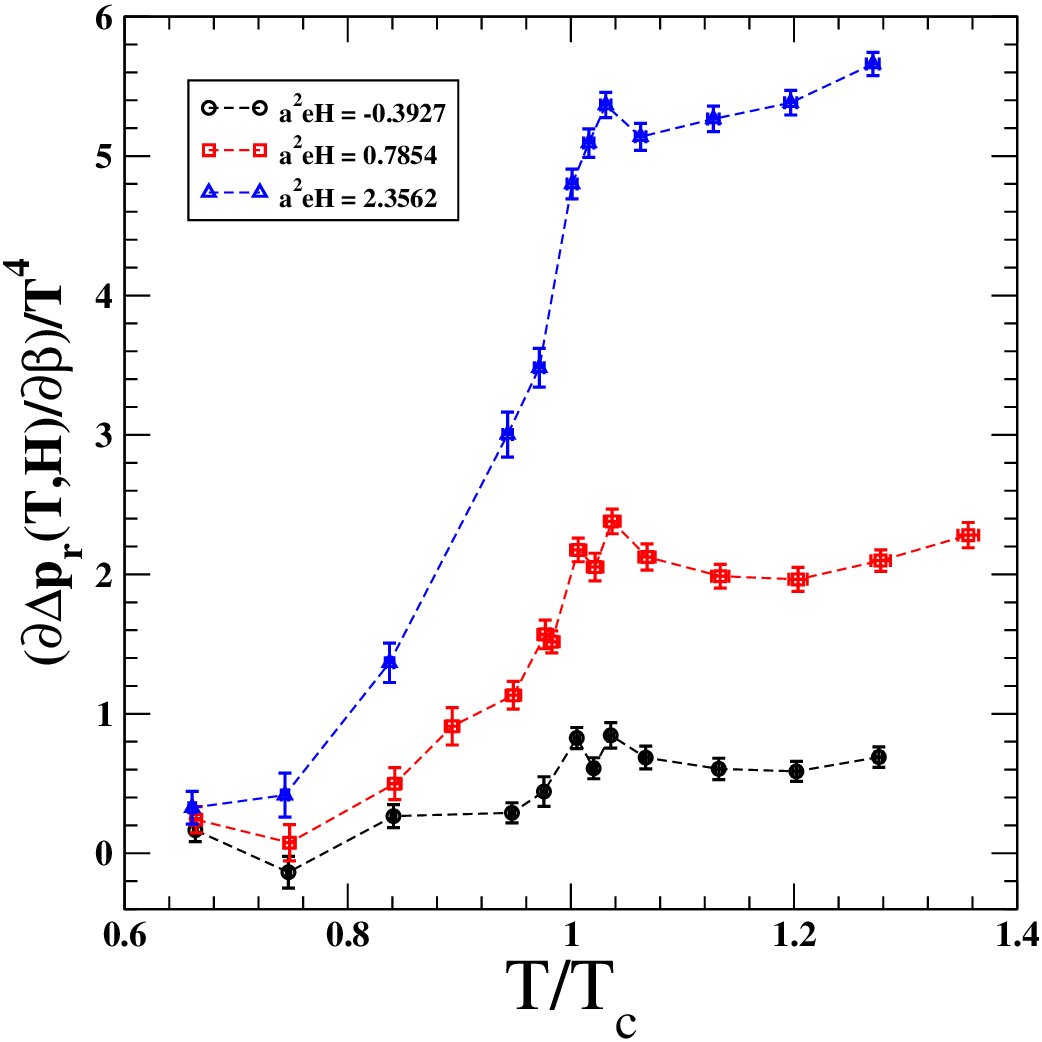} 
\caption{ $ \frac{1}{T^4} \frac{\partial\Delta \, p_r(T,H)}{\partial \, \beta}  $,  Eq.~(\protect\ref{4.9}),
 versus $T/T_c$  for    $q_f = \frac{2}{3}$  $n_{\text{ext}} =  \, 1 \, , \, 3$, 
 and   $q_f =  - \frac{1}{3}$   $n_{\text{ext}} =  \, 1$.}
\label{pressure_derivative}
\end{figure}
\begin{figure}[th] 
\centering
\includegraphics[width=0.7\textwidth]{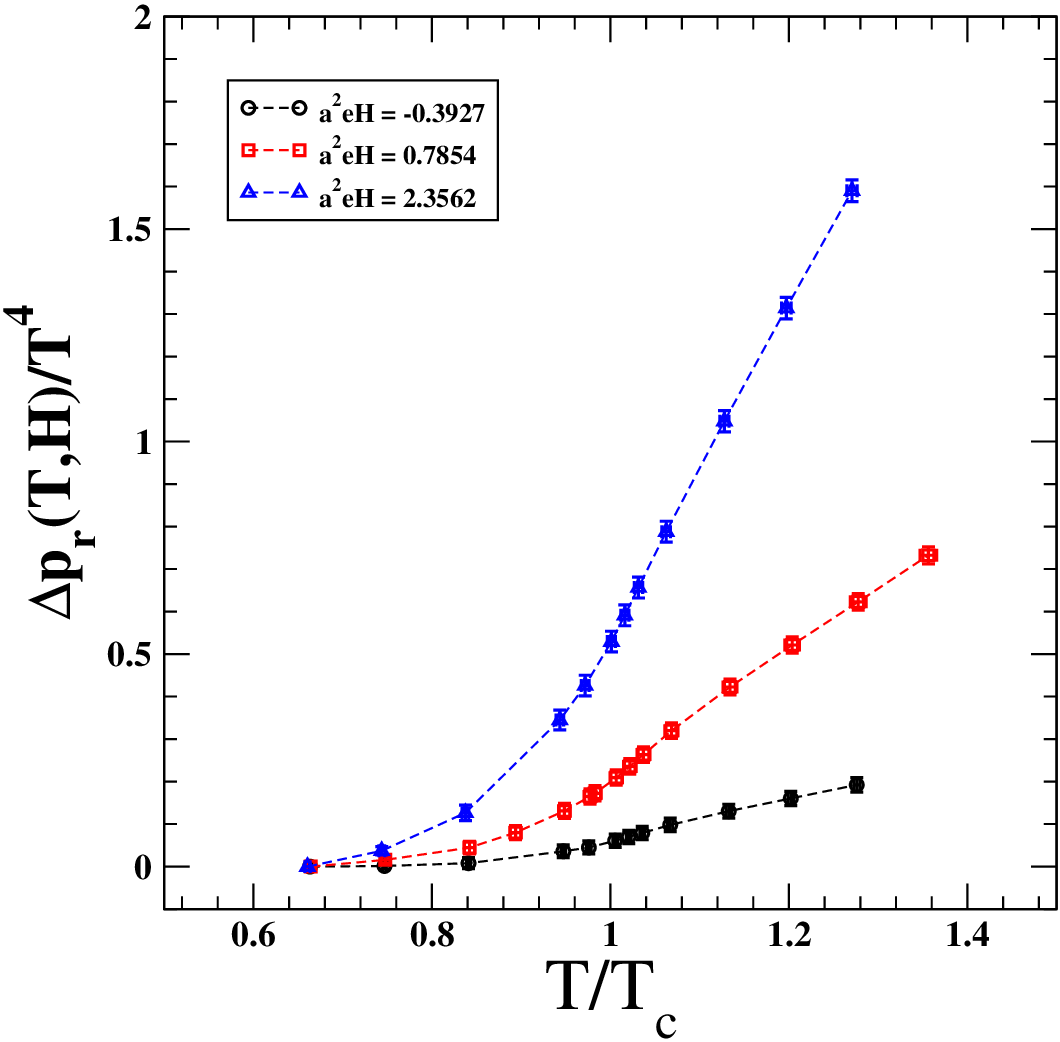} 
\caption{ $ \frac{\Delta \, p_r(T,H)}{T^4} $  
 versus $T/T_c$  for   $q_f = \frac{2}{3}$  $n_{\text{ext}} =  \, 1 \, , \, 3$, 
 and   $q_f =  - \frac{1}{3}$   $n_{\text{ext}} =  \, 1$.}
\label{pressure}
\end{figure}
\begin{figure}[th] 
\centering
\includegraphics[width=0.7\textwidth]{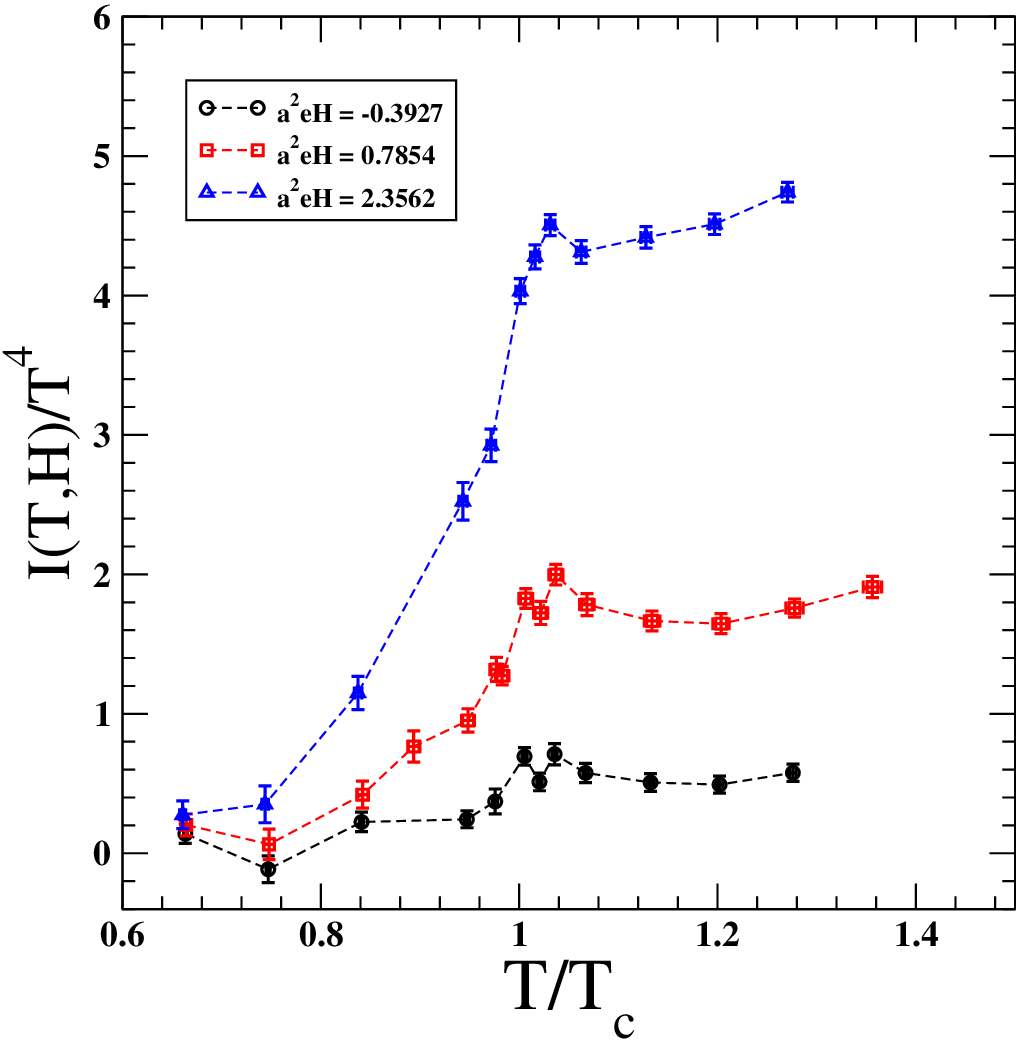} 
\caption{ The interaction measure $ \frac{I(T,H)}{T^4} $,  Eq.~(\protect\ref{4.10}),
 versus $T/T_c$  for    $q_f = \frac{2}{3}$  $n_{\text{ext}} =  \, 1 \, , \, 3$, 
 and   $q_f =  - \frac{1}{3}$   $n_{\text{ext}} =  \, 1$.}
\label{interaction_measure}
\end{figure}
\begin{figure}[th] 
\centering
\includegraphics[width=0.7\textwidth]{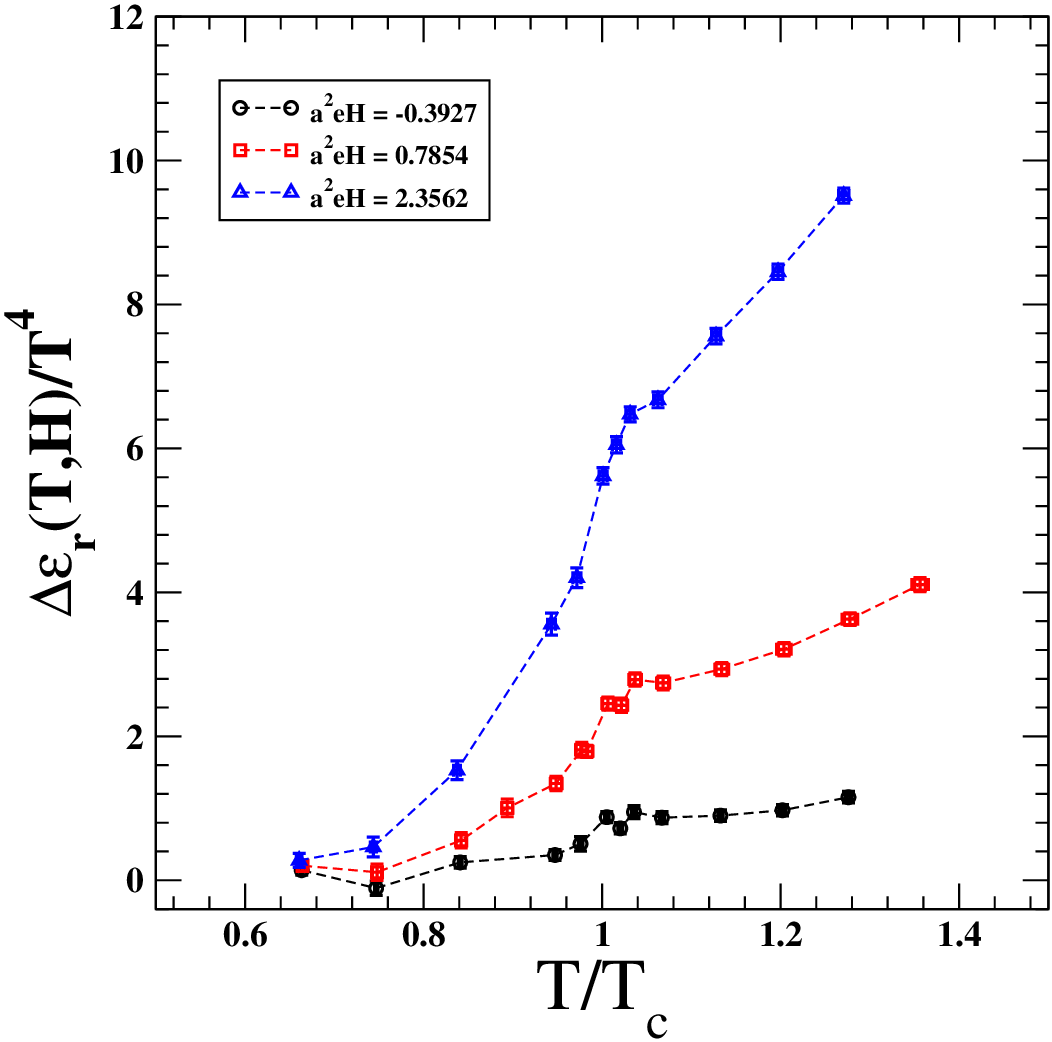} 
\caption{ $  \frac{\Delta \varepsilon_r(T,H)}{T^4}  $ , Eq.~(\protect\ref{4.12}),
 versus $T/T_c$  for    $q_f = \frac{2}{3}$  $n_{\text{ext}} =  \, 1 \, , \, 3$, 
 and   $q_f =  - \frac{1}{3}$   $n_{\text{ext}} =  \, 1$.}
\label{energydensity}
\end{figure}
As usual, we need to renormalize the free energy density by subtracting   the divergent zero-point energy. To do this it is enough
to subtract the zero temperature contribution. Thus we define:
\begin{equation}
\label{4.7}
f_r(T,H) \;  = \;    f(T,H)  \; - \;    f(0,H) \; \; , \; \;   p_r(T,H) \; = \; - \;   f_r(T,H) \; .
\end{equation}
The zero temperature contributions are conventionally obtained  by performing simulations
on lattices with $L_t = L_s$. Moreover, since we are interested in the thermal magnetic
properties of our system, we will focus on:
\begin{equation}
\label{4.8}
\Delta \, f_r(T,H) \;  \equiv \;    f_r(T,H)  \; - \;    f_r(T,H=0) \; \; , \; \;   \Delta \, p_r(T,H) \; = \; - \;  \Delta \, f_r(T,H) \; .
\end{equation}
In a Monte Carlo simulation, one cannot compute the partition function directly.
The most frequently used method in practice is the integral method, in which a derivative of the
free energy with respect to some parameter serves as observable, which then gets integrated again to
yield the free energy density. Since we are doing simulations at fixed $L_t$ , it is convenient to take derivatives with respect to
the bare gauge coupling $\beta$. The expectation values of the derivatives with respect to $\beta$  of our partition function correspond to
the average Wilson action. Thus, we have:
\begin{equation}
\label{4.9}
 \frac{1}{T^4} \frac{\partial\Delta \, p_r(T,H)}{\partial \, \beta}   \,  = \,  - \;  \frac{L_t^3}{L_s^3} 
\left \{ \left ( \langle S_W  \rangle_{H,T}  - \langle S_W \rangle_{H,0}  \right ) \; - \; 
\left ( \langle S_W  \rangle_{H=0,T}  - \langle S_W \rangle_{H=0,0}  \right )   \right \} \; .
\end{equation}
In Fig.~\ref{pressure_derivative}  we report our results for  the $\beta$-derivative of $\Delta \, p_r(T,H)$  (normalized to $T^4$)
versus the ratio $T/T_c$ for three different values of the magnetic field. We recall that  the temperature corresponding to a given value of the
gauge coupling is given by the relation $T= \frac{1}{L_t a(\beta)}$. For the dependence of the lattice spacing on the gauge
coupling we used the two-loop $\beta$-function. Accordingly, we have:
\begin{equation}
\label{4.9-bis}
 a(\beta) \;  \Lambda_{\mathrm{QCD}} \;  =  \;  f_{{\mathrm{QCD}}}(\beta) \; , 
\end{equation}
where $ f_{{\rm{QCD}}}(\beta)$ is the asymptotic scaling function of  QCD with one dynamical fermion $N_f = 1$:
\begin{equation}
\label{4.9-tris}
f_{{\rm{QCD}}}(\beta) = \left( \frac{6 b_0}{\beta} \right)^{- b_1/(2b_0^2)} 
\, \exp \left( - \frac{\beta}{12 b_0}\right)  \, ,  \; 
 b_0  =  \frac{11 - \frac{2}{3} N_f}{(4\pi)^2} \; \; , \; \; b_1  =  \frac{102 - \frac{38}{3} N_f}{(4\pi)^4}  \; .
\end{equation}
For definitiveness, the critical temperatures have been obtained by using the pseudocritical gauge coupling estimated
by means of the chiral susceptibilities.  Obviously, a direct determination of the physical scale  should be preferable.
 However, it is known that the lattice violations to the asymptotic
scaling law  Eq.~(\ref{4.9-bis}) are within a few percent. So that, the adopted approximation is adequate to the purpose of the present
 exploratory study. \\
From the derivative of the pressure, after numerical integration, we  may easily obtain   $\Delta \, p_r(T,H)  = -  \Delta \, f_r(T,H)$.
In Fig.~\ref{pressure} we display the normalized pressure $\Delta \, p_r(T,H)$ versus the temperature
 for  three different values of the magnetic field.  
Fig.~\ref{pressure} shows  that the magnetic contributions to the renormalized pressure is clearly different from
zero even for $T < T_c$, and it seems to vanish rapidly for low temperatures. This behavior can be naturally 
accounted for within the  Hadron Resonance Gas model (see, for instance, Ref.~\cite{Endrodi:2013cs,Kamikado:2014bua}).
On the other hand, for temperatures above the critical temperature the pressure increases rapidly in
qualitative agreements with  perturbative calculations in the high-temperature regime~\cite{Elmfors:1993wj}. \\
As concern the energy density, using  Eq.~(\ref{4.3}) we may write:
\begin{equation}
\label{4.10}
 \frac{I(T,H)}{T^4} \; = \;   \frac{\Delta \varepsilon_r(T,H) \, - \, 3 \, \Delta p_r(T,H)}{T^4} \; =  \;  T \;  \frac{d \, \beta }{d \, T} \; 
  \frac{\partial }{\partial \, \beta}  \left [ \frac{\Delta \, p_r(T,H)}{T^4} \right ] \; ,
\end{equation}
\begin{equation}
\label{4.11}
 T \;  \frac{d \, \beta }{d \, T}  \;  = \;  - \;  a \;  \frac{d \, \beta }{d \, a}  \;  \equiv  \; R_{\beta}(\beta)  \; ,
\end{equation}
where $ R_{\beta}(\beta)$ is the lattice $\beta_{QCD}$-function.
According to our approximation, we have:
\begin{equation}
\label{4.11bis}
%
% R_{\beta}(\beta)  \;   = \;  - \; \frac{  \frac{d \,  }{d  \beta}  f_{{\rm{QCD}}}(\beta)}{ f_{{\rm{QCD}}}(\beta)}  \; .
R_{\beta}(\beta)  \;   = \;  - \; \frac{ f_{{\rm{QCD}}}(\beta)}{  \frac{d \,  }{d  \beta}  f_{{\rm{QCD}}}(\beta)}  \; .
\end{equation}
Using  Eqs.~(\ref{4.10}),   (\ref{4.11}), and   (\ref{4.11bis})  we determined the so-called
interaction measure  displayed in   Fig.~\ref{interaction_measure}  for three different values of the magnetic field. 
\begin{figure}[ht] 
\centering
\includegraphics[width=0.7\textwidth]{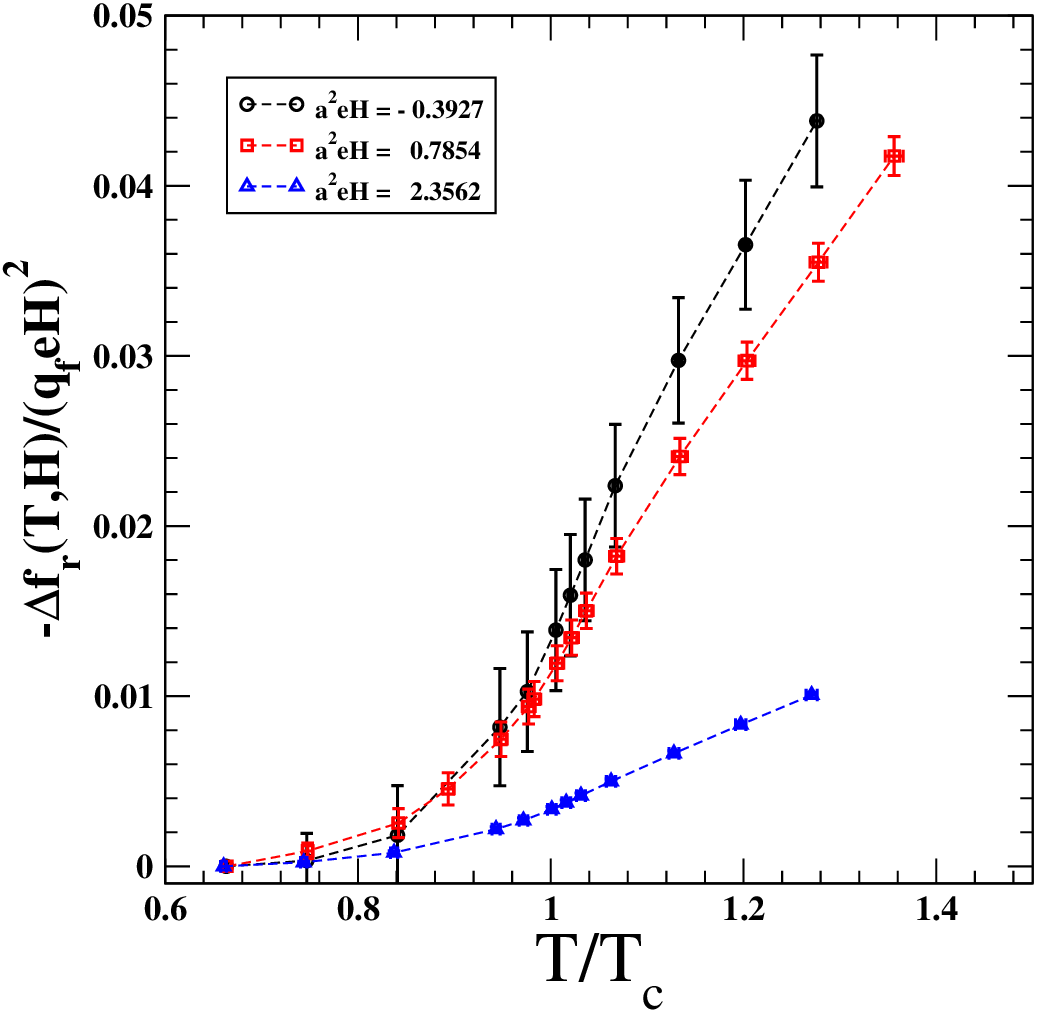} 
\caption{ $ \frac{\Delta \, f_r(T,H)}{(q_f eH)^2} $   
 versus $T/T_c$  for  $q_f =  - \frac{1}{3}$   $n_{\text{ext}} =  \, 1$, and
 $q_f = \frac{2}{3}$  $n_{\text{ext}} =  \, 1 \, , \, 3$.}
\label{freeenergydensity}
\end{figure}
\begin{figure}[ht] 
\centering
\includegraphics[width=0.7\textwidth]{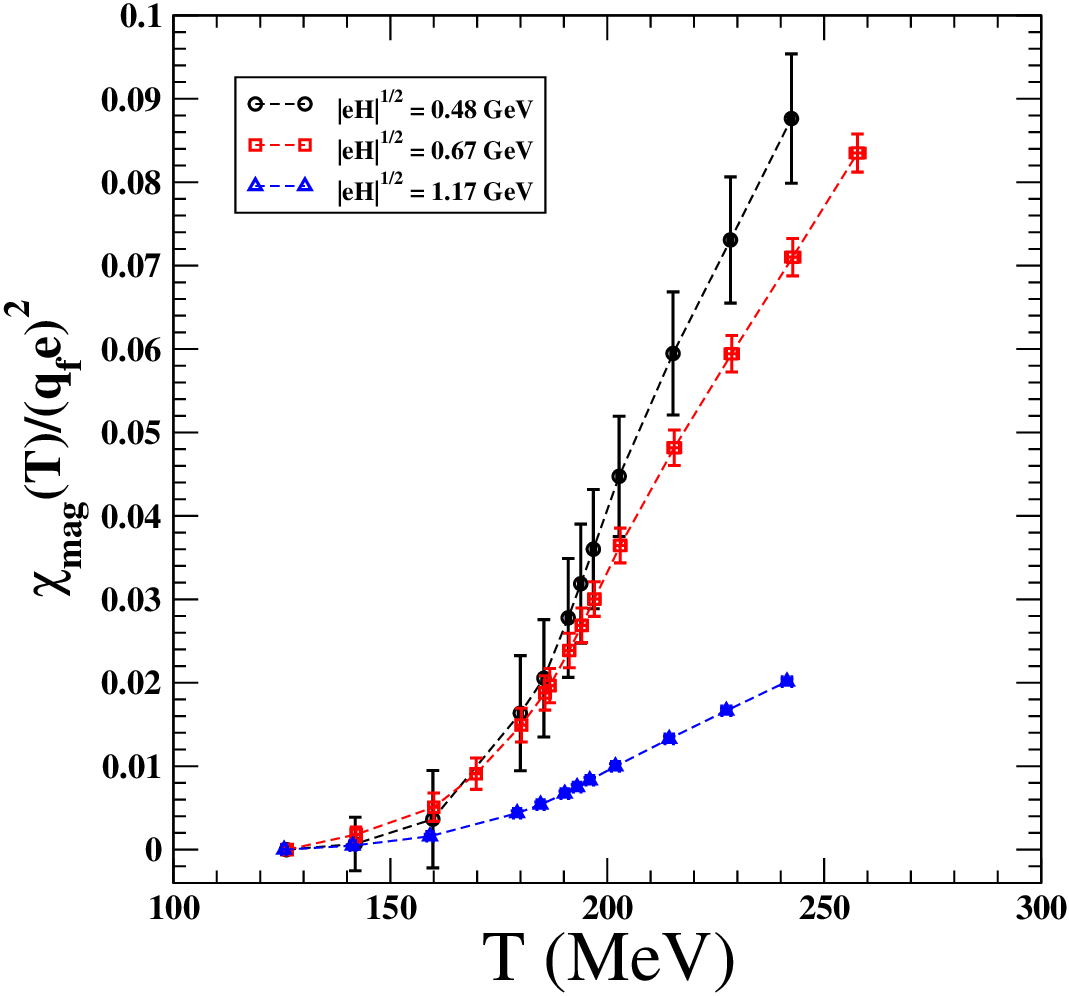} 
\caption{ The magnetic susceptibility $\frac {\chi}{q_f^2 e^2} $   
 as a function of the temperature for different magnetic field strengths.}
\label{susceptibility}
\end{figure}

After that,  the magnetic contributions to the renormalized energy density can be straightforwardly obtained as:
\begin{equation}
\label{4.12}
\frac{\Delta \varepsilon_r(T,H)}{T^4} \; =  \;  \frac{\Delta I(T,H)}{T^4} \; + \;  3 \, \frac{\Delta p_r(T,H)}{T^4}  \; .
\end{equation}
In  Fig.~\ref{energydensity} we show the renormalized energy density.  Even for the  magnetic contribution to
the energy density we find two different regimes for $T < T_c$ (confined phase) and $T > T_c$ (deconfined phase).
Having determined the magnetic contribution to the pressure and energy density, in principle
one can construct the equation of state and obtain the entropy density and the  speed of sound by means of   Eqs.~(\ref{4.5}) and  (\ref{4.6}). 
For the purposes of the present paper we do not discuss any further this matter. We, merely, observe that in the deconfined phase  $T > T_c$
the magnetic contribute to the energy density increases  slower with respect to the pressure by increasing the temperature.  This behavior leads
to a stiffening of the equation of state. \\
Let us, finally, address the problem of the magnetic susceptibility.  As is well known,
for small magnetic field strengths we may write for the free energy density (see, for instance, Ref.~\cite{Landau:1984}):
\begin{equation}
\label{4.13}
  \Delta \, f_r(T,H) \,  = \,  -  \, \frac{1}{2} \; \chi_{mag}(T) \; H^2  \; ,
\end{equation}
where $ \chi_{mag}$ is the magnetic susceptibility.
Therefore, to determine the magnetic susceptibility we need to check if   $ \Delta \, f_r(T,H) =  - \Delta \, p_r(T,H)$ 
scales with $H^2$  at least for small  enough magnetic field strengths. To this end,  in Fig~\ref{freeenergydensity} we display  
 $\frac{\Delta \, f_r(T,H)}{(q_f eH)^2}$ for different values of the magnetic field strength.   
 In fact, we see that the free energy density seems to scale with $H^2$ for  ($q_f = \frac{2}{3}$, $n_{\text{ext}} =  \, 1 $), 
 and   ($q_f =  - \frac{1}{3}$,   $n_{\text{ext}} =  \, 1$), and for temperatures not too far from the critical temperature.
 We note, however, that  in the high-temperature region  physical observables are more affected by finite volume and cutoff effects.
On the other hand, we see clearly that for the strongest
 magnetic field used in this paper   ($q_f = \frac{2}{3}$,     $n_{\text{ext}} =  \, 3$)  the scaling  with $H^2$ is badly violated.
It is useful to  give the corresponding values of the magnetic field in physical units. To this end,
we use the  known flavor dependence of the QCD critical temperature reported in Ref.~\cite{Karsch:2000kv} to infer
that for $N_f=1$ the critical temperature is $T_c \sim 190 \, {\mathrm{MeV}}$.  This corresponds to a lattice spacing $a \simeq  0.26 \, {\mathrm{fm}}$.
So that, in the critical region, we estimate $\sqrt{|e H|} \simeq 0.48 , \, 0.67 , \, 1.17 \, {\mathrm{GeV}}$ corresponding to   
$n_{\text{ext}} =  \, -0.5 \, , \, 1 \, , \, 3$  respectively. 
Thus, we see that for magnetic field strengths not exceeding $1.0 \, GeV$ the free energy density seems to display
an approximate scaling with $H^2$ within our statistical uncertainties. 
This is in qualitative agreement with the results in Ref.~\cite{D'Elia:2010nq} where the strongest magnetic field 
used was   $\sqrt{|e H|} \simeq 0.85 \, {\mathrm{GeV}}$. In any case, for physical applications, 
we recall that the magnetic fields relevant for heavy-ion collision experiments
are  of order  $\sqrt{|e H|} \sim 0.1 \, {\mathrm{GeV}}$. \\
To determine the magnetic susceptibility we are lead to consider the free energy density for  $\sqrt{|e H|} \leq 1.0 \, {\mathrm{GeV}}$ where
we can safely apply Eq.~(\ref{4.13}). 
Indeed, in  Fig.~\ref{susceptibility}   we report our determination of the magnetic susceptibility as  function of the temperature. 
For comparison, we also display our determination of the magnetic susceptibility for   $\sqrt{|e H|} \simeq 1.17 \, {\mathrm{GeV}}$.
From  Fig.~\ref{susceptibility}  we see that the magnetic susceptibility is positive in the whole temperature range explored
in the present study. Moreover, the magnetic susceptibility increases monotonically with the temperature. Therefore,
the strongly interacting medium behaves as a paramagnetic substance both below and above the critical temperature $T_c$.
It is remarkable that our results for the magnetic susceptibility is in fair qualitative and quantitative agreement with
Ref.~\cite{Bonati:2013vba}  where it has been considered $N_f = 2+1$ QCD with physical quark masses, discretized on
a lattice by stout improved staggered fermions and a tree level improved Symanzik pure gauge action.
\section{Conclusions and Discussion}
\label{sec5}
In conclusion, let us summarize briefly the main results of the present paper. We investigated QCD  with one flavor of 
staggered quark in an external magnetic field on the lattice.   The external magnetic field 
has been introduced by means of the so-called Schr\"odinger functional. We have investigated the magnetic properties
of one-flavour quarks and gluons in thermal equilibrium for magnetic field strengths up to  $\sqrt{|e H|} \leq 1.17 \, {\mathrm{GeV}}$.
In particular, we focused on the effects of the magnetic field on several local observables and found results in qualitative
agreement with recent results in the literature obtained with a different method, as described in Sect.~\ref{sec1}, to implement
external magnetic fields in QCD on the lattice. We have, also, looked for the effects of the magnetic field on the critical
temperature. Surprisingly, we found that the critical temperature does not change even for the strongest magnetic
field used in the present work. This is in striking contrast with the results in the literature. However, since we used
one flavor rooted staggered quark  which is known to be strongly affected by taste symmetry violation effects, one could
suspect that our results on the critical temperature is merely due to lattice artifacts. Indeed, presently we are
simulating the same physical system by adopting  highly improved staggered quarks (HISQ) where 
 the  taste symmetry violations are dramatically reduced. Nevertheless,   our preliminary simulations adopting  HISQ quarks
 to do not yet display  a clear dependence of the pseudocritical temperature on the background magnetic field. 
 In any event,  we plan to report progress on this subject in a future paper. \\
 We evaluated the magnetic contributions to the pressure, energy density, and free energy. Our results
 are in qualitative agreement with previous investigations. In particular, we confirm that the free energy density
 scales with $H^2$ for small enough magnetic field strengths. Moreover, we determined the magnetic susceptibility 
 and found that the strongly interacting medium behaves like a paramagnetic substance both below and above the critical temperature 
in agreement with  previous results in the literature. 

\acknowledgments
This work was in part based on the MILC collaboration public lattice gauge theory code,
see \url{http://physics.utah.edu/~detar/milc.html}. This work was partially supported by the INFN SUMA project.
Simulations have been performed on Blue Gene/Q Fermi and Galileo Supercomputer
Cluster at CINECA (CINECA-INFN agreement), on the BC2S cluster in Bari,
and on the Zefiro cluster in Pisa.

\bibliography{qcd}
\providecommand{\href}[2]{#2}\begingroup\raggedright\endgroup

\end{document}